# Electromagnetic fields in a time-varying medium: Exceptional points and operator symmetries

Theodoros T. Koutserimpas, *Student Member, IEEE*, Romain Fleury, *Member, IEEE*

*Abstract*— In this paper, we study the interactions of electromagnetic waves with a non-dispersive dynamic medium that is temporally dependent. Electromagnetic fields under material time-modulation conserve their momentum but not their energy. We assume a time-variation of the permittivity, permeability and conductivity and derive the appropriate time-domain solutions based on the causality state at a past observation time. We formulate a time-transitioning state matrix and connect the unusual energy transitions of electromagnetic fields in time-varying media with the exceptional point theory. This state-matrix approach allows us to analyze further the electromagnetic waves in terms of parity and time-reversal symmetries and signify parity-time symmetric wave-states without the presence of a spatially symmetric distribution of gain and loss, or any inhomogeneities and material periodicity. This paper provides a useful arsenal to study electromagnetic wave phenomena under time-varying media and points out novel physical insights connecting the resulting energy transitions and electromagnetic modes with exceptional point physics and operator symmetries.

*Index Terms*— electromagnetic propagation, exceptional point theory, time-domain analysis, time-varying systems, time-reversal, parity operator symmetry.

## I. INTRODUCTION

ELECTROMAGNETIC wave propagation under time-modulation has been first studied by Morgenthaler [1]. It was showed that electromagnetic waves reflect not only due to spatial inhomogeneities, but also due to temporal jumps of the material parameters ($\varepsilon, \mu$). Later studies proceeded in the integral representation of some examples with analytical solutions showing interesting electromagnetic radiation phenomena, i.e. longitudinal far field propagation, radiation by steady currents, etc. [2]–[5]. In addition, transmission and reflection studies have been made for time-periodic media showing interesting phenomena; in particular parametric amplification and frequency variance, by utilizing harmonic balance techniques or the solutions of the corresponding hypergeometric functions [6]–[13]. Time-modulation effects on transmission lines have also been investigated, indicating the possibility of perfect matching scenarios and wave amplification [14], [15].

In this paper, we expand the existing literature on the interaction of electromagnetic waves with time-varying media by considering a general medium time-variation and providing a mathematical formulation of the electromagnetic field solutions using the separation of variables technique. We utilize the momentum-Fourier integral transforms and deliver time-domain solutions, without presuming a certain temporal profile for the wave parameters. The determination of specific solutions requires the knowledge of the causality of the fields. Their derivation is found as long as the fields are known for a given past observation time. Furthermore, we derive the appropriate scattering matrices, which deal with temporal material discontinuities.

In the last part of this article, we provide a modern insight to the exhibited electromagnetic phenomena, by connecting the uncommon energy transitions that transpire from the time-dynamic wave and material interactions with the exceptional point theory in the temporal problem of a step variation of the medium (as was first established by Morgenthaler in [1]). Such analysis is done by formulating the time-transitioning state matrix and derive the proper stability conditions, which determine stable wave-states or parametrically amplifying and evanescent ones, establishing the importance of the observation time of the fields in relation with their energy. Finally, we investigate the parity and time-reversal symmetries and report the possibility to induce parity-time wave states for time-symmetric modulations without the presence of material gain and loss [16], [17] nor assuming a time-dependent permittivity inspired by $\mathcal{PT}$ – symmetry [18].

## II. ELECTRIC DISPLACEMENT AND MAGNETIC INDUCTION ANALYSIS

We assume the case of electromagnetic wave propagation in a homogeneous time-varying medium. The medium's properties are defined by temporally dependent permittivity, conductivity and permeability: $\varepsilon = \varepsilon(t)$, $\sigma = \sigma(t)$ and $\mu = \mu(t)$. Such medium has been first considered by Morgenthaler (excluding the conductivity) in [1]. Maxwell's equations are: $\nabla \times \mathbf{H} = \mathbf{J} + \partial_t \mathbf{D} + \sigma \mathbf{E}$, $\nabla \times \mathbf{E} = -\partial_t \mathbf{B}$, $\nabla \cdot \mathbf{D} = \rho$ and $\nabla \cdot \mathbf{B} = 0$ while the constitutive relations of the

This work was supported by the Swiss National Science Foundation under Grant 172487.
The authors are with the Laboratory of Wave Engineering, Swiss Federal Institute of Technology in Lausanne, 1015 Lausanne, Switzerland.

medium are assumed: $\mathbf{D} = \varepsilon \mathbf{E}$ and $\mathbf{B} = \mu \mathbf{H}$, which imply an instantaneously responding medium (for a thorough analysis regarding the wave parameters of a dispersive but time-varying medium the interested reader may be referred to [19]).

The resulting second order differential equations are (derived in a similar way as the derivations of inhomogeneous stationary media in [20]).

$$\nabla \times \nabla \times \mathbf{E} + \frac{\partial}{\partial t}\left[\mu(t)\frac{\partial(\varepsilon(t)\mathbf{E})}{\partial t}\right] + \frac{\partial(\mu(t)\sigma(t)\mathbf{E})}{\partial t} = -\frac{\partial(\mu(t)\mathbf{J})}{\partial t}, \quad (1)$$

$$\nabla \times \nabla \times \mathbf{H} + \frac{\partial}{\partial t}\left[\varepsilon(t)\frac{\partial(\mu(t)\mathbf{H})}{\partial t}\right] + \sigma(t)\frac{\partial(\mu(t)\mathbf{H})}{\partial t} = \nabla \times \mathbf{J}. \quad (2)$$

(For the more general case which includes inhomogeneities see Appendix A). For a stationary problem, inhomogeneities result overtly in transmitted and reflected fields [21]. The wave frequency is preserved whereas the wave vector changes. On the other hand, in the case of a time-dynamic homogeneous medium, which is derived in Eqs. (1) and (2), the wave vector is invariant, while the operating frequency of the wave is altered (the phenomenon is often called adiabatic wavelength conversion) [1]. (For an alternative proof based on the Poynting and the momentum conservation theorems see Appendix B). It is important to mention that these overwhelming wave-reconfiguration abilities come with a major cost, namely energy consumption. Such artificial media have to be externally supplied by energy to effectively change properties and should thus be considered active, depending on the modulation of the medium parameters (Appendix B). Although these media should be considered as active, they do not necessarily provide amplification to the electromagnetic fields.

The mathematical modeling of such equations is easier handled, if we consider the electric displacement and the magnetic induction fields, which are required to be continuous under the time-variations of the medium [1]. The corresponding equations for the electric displacement and magnetic induction fields are:

$$\nabla \times \nabla \times \mathbf{D} + \varepsilon(t)\frac{\partial}{\partial t}\left[\mu(t)\frac{\partial \mathbf{D}}{\partial t}\right]$$
$$+\varepsilon(t)\frac{\partial}{\partial t}\left[\mu(t)\sigma(t)\varepsilon^{-1}(t)\mathbf{D}\right] = -\varepsilon(t)\frac{\partial(\mu(t)\mathbf{J})}{\partial t}, \quad (3)$$

$$\nabla \times \nabla \times \mathbf{B} + \mu(t)\frac{\partial}{\partial t}\left[\varepsilon(t)\frac{\partial \mathbf{B}}{\partial t}\right]$$
$$+\mu(t)\sigma(t)\frac{\partial \mathbf{B}}{\partial t} = \mu(t)\nabla \times \mathbf{J}. \quad (4)$$

An extension of the potential theory for time-varying media, with an appropriate Lorentz gauge condition is shown in Appendix C. The radiation by currents, which satisfy Eqs. (3) and (4) for isotropic time-dynamic media has been studied extensively by Budko [2] using the volume integral equation method in combination with an appropriate variable transformation of the time variable $t$ for specific functions of $\varepsilon, \mu$. It was found that by taking into account the time-dynamics and causality, a time-varying medium can permit propagation of longitudinal electromagnetic waves, and furthermore can radiate waves from steady currents, as the electromagnetic disturbances from the sudden excitation of the currents continue to sway due to the time-dynamics of the medium.

For source-free regions Eqs. (3) and (4) have a zero right-hand side and become:

$$\nabla^2 \mathbf{D} - \varepsilon(t)\frac{\partial}{\partial t}\left[\mu(t)\frac{\partial \mathbf{D}}{\partial t}\right] - \varepsilon(t)\frac{\partial}{\partial t}\left[\mu(t)\sigma(t)\varepsilon^{-1}(t)\mathbf{D}\right] = 0, \quad (5)$$

$$\nabla^2 \mathbf{B} - \mu(t)\frac{\partial}{\partial t}\left[\varepsilon(t)\frac{\partial \mathbf{B}}{\partial t}\right] - \mu(t)\sigma(t)\frac{\partial \mathbf{B}}{\partial t} = 0. \quad (6)$$

We assume a solution for the magnetic induction or the electric displacement field $\mathbf{U} = \mathbf{B}, \mathbf{D}$ of the form: $\mathbf{U}(\mathbf{r},t) = \mathbf{R}(\mathbf{r})T(t)$, since Eqs. (5) and (6) are fully separable [6]. The separation of variables method allows both real and complex solutions for $\mathbf{R}(\mathbf{r})$ and $T(t)$. Real solutions of $\mathbf{R}(\mathbf{r})$ and $T(t)$ represent a summation of standing waves, whereas complex solutions represent a summation of propagating waves (both solutions are valid). We proceed below by finding the propagating-wave solutions (complex functions of $\mathbf{R}(\mathbf{r})$ and $T(t)$) for which the actual field in time-domain is: $\mathrm{Re}[\mathbf{U}(\mathbf{r},t)] = \mathrm{Re}[\mathbf{R}(\mathbf{r})T(t)]$.

The differential equation of $\mathbf{R}(\mathbf{r})$ can be easily shown to be the Helmholtz equation: $\nabla^2 \mathbf{R} = -k^2 \mathbf{R}$, where $k$ is the wavenumber. The wavenumber remains constant despite any material temporal change as indicated by the separability of the equations [1]. The complex solution of $\mathbf{R}(\mathbf{r})$, which is far away from the sources has the form for a given $k$ (imposed by the sources):

$$\mathbf{R}(\mathbf{r}) = \int_0^\pi \int_0^{2\pi} \mathbf{F}(\varphi,\theta)e^{i\mathbf{k}\cdot\mathbf{r}}d\varphi d\theta, \quad (7)$$

where $\mathbf{F}$ is a vector dependent on the spherical coordinates $\theta, \varphi$ and $\mathbf{k} = k\sin\theta\cos\varphi \cdot \hat{x}_1 + k\sin\theta\sin\varphi \cdot \hat{x}_2 + k\cos\theta \cdot \hat{x}_3$ which is the wavevector. We assume that for a substantial period of time there was no conductivity, and the field solutions filled the whole region. Such solutions correspond to the source-free modes of Maxwell's equations in time-varying media. (An infinite number of wavenumbers can be propagating). On the other hand, the solution of the time-function is strongly dependent on the type of the time-variation of the medium (see Appendix D). It is though obvious that the resulting second order differential equations result in two independent solutions: $T_1(k,t)$ and $T_2(k,t)$ (which we assume complex as we formulate the system using propagating wave-bases).

For slowly and continuously varying material-parameters the Liouville-Green approximation [22] can be directly applied and gives us an estimation of the fields (see Appendix D):

$$T_b(k,t) \approx \sqrt[4]{\frac{\mu}{\varepsilon}}e^{-\frac{1}{2}\int\frac{\sigma}{\varepsilon}dt}\left(C_1^b e^{-ik\int\frac{dt}{\sqrt{\varepsilon\mu}}} + C_2^b e^{+ik\int\frac{dt}{\sqrt{\varepsilon\mu}}}\right), \quad (8)$$

$$T_d(k,t) \approx e^{-\int\frac{\sigma}{\varepsilon}dt}\int e^{\frac{1}{2}\int\frac{\sigma}{\varepsilon}dt}\sqrt[4]{\frac{\mu}{\varepsilon}}\left(C_1^d e^{-ik\int\frac{dt}{\sqrt{\varepsilon\mu}}} + C_2^d e^{+ik\int\frac{dt}{\sqrt{\varepsilon\mu}}}\right)dt, \quad (9)$$

where $T_b(k,t)$ is the solution for the magnetic induction,

$T_d(k,t)$ is the solution for the electric displacement, $C_1(k)$ and $C_2(k)$ represent the coefficients of the electromagnetic modes which propagate at the positive and negative direction, respectively. The presence of the conductivity results in the attenuation of the fields by a $\exp\left(-\frac{1}{2}\int \frac{\sigma}{\varepsilon} dt\right)$ factor, while the velocity is $c_{\text{eff}}(t) = \frac{\int dt/\sqrt{\varepsilon(t)\mu(t)}}{t}$.

The general solution, if $T(k,t)$ and $\mathbf{F}(\varphi,\theta)$ are known, is:

$$\mathbf{U}(\mathbf{r},t) = \int \left(\mathbf{F}_1(\mathbf{k})T_1(k,t) + \mathbf{F}_2(\mathbf{k})T_2(k,t)\right) e^{i\mathbf{k}\cdot\mathbf{r}} d^3k, \quad (10)$$

where $\mathbf{F}_1(\mathbf{k}) = C_1(k)\mathbf{F}(\varphi,\theta)$, $\mathbf{F}_2(\mathbf{k}) = C_2(k)\mathbf{F}(\varphi,\theta)$ and $C_{1,2}(k)$ are the coefficients corresponding to the time-function $T_{1,2}(k,t)$ solutions. Obviously, the time-domain solution is $\text{Re}[\mathbf{U}(\mathbf{r},t)]$.

### A. Causality of the solutions: Determination of $\mathbf{F}_{1,2}(\mathbf{k})$ by the observation of the field at $t = \tau$

At $t = \tau$ we measure $\mathbf{U}(\mathbf{r},\tau)$ and $\partial_t \mathbf{U}(\mathbf{r},\tau)$. The time-domain observation of the fields gives us real values of $\mathbf{E}(\mathbf{r},\tau)$ and $\mathbf{H}(\mathbf{r},\tau)$ [which are functions of $\mathbf{r} = (x_1, x_2, x_3)$]. In order to apply them as temporal boundaries for the complex (propagating) $\mathbf{U}$ we shall convert them in the complex space. This is feasible by the direct application of the Hilbert transform, which leads to the analytic form of the signals. Precisely: if $\boldsymbol{\psi}(\mathbf{r})$ is $\mathbf{E}(\mathbf{r},\tau)$ or $\mathbf{H}(\mathbf{r},\tau)$, then its analytic function is $\tilde{\boldsymbol{\psi}}(\mathbf{r}) = \prod_{j=1}^{3}\left[\delta(x_j) + \frac{i}{\pi x_j}\right] *** \boldsymbol{\psi}(\mathbf{r})$, where $\delta(x_j)$ is the delta function with respect to the Cartesian coordinate $x_j$ and $***$ is the 3-fold convolution. For $\mathbf{U}(\mathbf{r},\tau)$: $\tilde{\mathbf{D}}(\mathbf{r},\tau) = \varepsilon(\tau)\tilde{\mathbf{E}}(\mathbf{r},\tau)$ or $\tilde{\mathbf{B}}(\mathbf{r},\tau) = \mu(\tau)\tilde{\mathbf{H}}(\mathbf{r},\tau)$, and for $\partial_t \mathbf{U}(\mathbf{r},\tau)$: $\partial_t \tilde{\mathbf{D}}(\mathbf{r},\tau) = \nabla \times \tilde{\mathbf{H}}(\mathbf{r},\tau) - \sigma(\tau)\tilde{\mathbf{E}}(\mathbf{r},\tau)$ or $\partial_t \tilde{\mathbf{B}}(\mathbf{r},\tau) = -\nabla \times \tilde{\mathbf{E}}(\mathbf{r},\tau)$.

The Wronski determinant $\Delta(t)$ has to be nonzero [23]. For the magnetic induction $\Delta_b(t) = \Delta_b(\tau)\frac{\varepsilon(\tau)}{\varepsilon(t)}\exp\left(-\int_\tau^t \frac{\sigma}{\varepsilon} dt\right)$ and for the electric displacement $\Delta_d(t) = \Delta_d(\tau)\frac{\mu(\tau)}{\mu(t)}\exp\left(-\int_\tau^t \frac{\sigma}{\varepsilon} dt\right)$.

The solution of the temporal boundary problem concludes with the determination of $\mathbf{F}_1(\mathbf{k})$ and $\mathbf{F}_2(\mathbf{k})$. We apply the inverse Fourier integral on $\mathbf{U}(\mathbf{r},\tau)$ and $\partial_t \mathbf{U}|_{t=\tau}$, which leads to the solutions:

$$\mathbf{F}_1(\mathbf{k}) = \frac{\frac{dT_2(k,\tau)}{dt}\int \mathbf{U}(\mathbf{r},\tau)e^{-i\mathbf{k}\cdot\mathbf{r}}d^3r - T_2(k,\tau)\int \frac{\partial \mathbf{U}(\mathbf{r},\tau)}{\partial t}e^{-i\mathbf{k}\cdot\mathbf{r}}d^3r}{8\pi^3 \Delta(\tau)}, \quad (11)$$

$$\mathbf{F}_2(\mathbf{k}) = -\frac{\frac{dT_1(k,\tau)}{dt}\int \mathbf{U}(\mathbf{r},\tau)e^{-i\mathbf{k}\cdot\mathbf{r}}d^3r - T_1(k,\tau)\int \frac{\partial \mathbf{U}(\mathbf{r},\tau)}{\partial t}e^{-i\mathbf{k}\cdot\mathbf{r}}d^3r}{8\pi^3 \Delta(\tau)}. \quad (12)$$

From the temporal conditions at $t = \tau$ the existing wavevectors are defined (which are the $k$-content of the fields).

### B. Temporal material discontinuities: Scattering matrix formulation

It is well-known [1], [24] that under temporal discontinuities waves exhibit reflection. Therefore, we introduce in this subsection the mathematical treatment to deal with them. In a hypothetical scenario, at a time $t = \tau$ the temporal functions of $\varepsilon(t), \mu(t), \sigma(t)$ or their derivatives exhibit a discontinuity. Then, by the observation of the differential equations (see Apendix D) it is clear that this discontinuity can be carried to $d^2T(k,t)/dt^2$, leaving $T(k,t)$ and $dT(k,t)/dt$ to be continuous. Such continuity condition is sufficient when either the permeability or the permittivity are continuous. For instance, if the permeability is continuous the electric displacement field and its time derivative are continuous, whereas if the permittivity and conductivity are continuous the magnetic induction field and its time derivative are continuous. (In other cases the time-function solutions may involve delta-like distributions). If the solution for $t < \tau$ is:

$$T_<(k,t) = C_1^< T_1^<(k,t) + C_2^< T_2^<(k,t), \quad (13)$$

and for $t > \tau$:

$$T_>(k,t) = C_1^> T_1^>(k,t) + C_2^> T_2^>(k,t), \quad (14)$$

then the continuity relations at $t = \tau$ give us the temporal scattering matrix:

$$\begin{pmatrix} C_1^> \\ C_2^> \end{pmatrix} = S_\tau \cdot \begin{pmatrix} C_1^< \\ C_2^< \end{pmatrix} =$$

$$\begin{pmatrix} \dfrac{\frac{dT_1^<(k,\tau)}{dt}T_2^>(k,\tau) - \frac{dT_2^>(k,\tau)}{dt}T_1^<(k,\tau)}{\frac{dT_1^>(k,\tau)}{dt}T_2^>(k,\tau) - \frac{dT_2^>(k,\tau)}{dt}T_1^>(k,\tau)}, & \dfrac{\frac{dT_2^<(k,\tau)}{dt}T_2^>(k,\tau) - \frac{dT_2^>(k,\tau)}{dt}T_2^<(k,\tau)}{\frac{dT_1^>(k,\tau)}{dt}T_2^>(k,\tau) - \frac{dT_2^>(k,\tau)}{dt}T_1^>(k,\tau)} \\ \dfrac{(T_1^<(k,\tau) - T_1^>(k,\tau))\left(\frac{dT_1^<(k,\tau)}{dt} - \frac{d\ln T_2^>(k,\tau)}{dt}T_1^<(k,\tau)\right)}{\frac{dT_1^>(k,\tau)}{dt}T_2^>(k,\tau) - \frac{dT_2^>(k,\tau)}{dt}T_1^>(k,\tau)}, & \dfrac{(T_2^<(k,\tau) - T_1^>(k,\tau))\left(\frac{dT_2^<(k,\tau)}{dt} - \frac{d\ln T_2^>(k,\tau)}{dt}T_2^<(k,\tau)\right)}{\frac{dT_1^>(k,\tau)}{dt}T_2^>(k,\tau) - \frac{dT_2^>(k,\tau)}{dt}T_1^>(k,\tau)} \end{pmatrix} \cdot \begin{pmatrix} C_1^< \\ C_2^< \end{pmatrix}, \quad (15)$$

where $S_{11}$ and $S_{22}$ are the parameters which correspond to the temporal transmission, whereas $S_{21}$ and $S_{12}$ are the parameters for the temporal reflection.

Let us assume the scenario that a wave propagates at a medium with $\varepsilon_1, \mu_1, \sigma_1$ (which are constant) and suddenly at $t = 0$ the medium changes its electromagnetic-material properties to $\varepsilon_2, \mu_2, \sigma_2$ (which are also constant). The ordinary differential equation that satisfies $T(k,t)$ for both **B** and **D** under constant values of $\varepsilon, \mu, \sigma$ is the same (see Appendix D). The time-domain solution as found by Maxwell's equations under nonstationary conditions (of either **B** and **D** depending on the continuity conditions) is of the form: $\zeta(\mathbf{r},t) = \zeta_0 \cos\left[\mathbf{k}\cdot\mathbf{r} - \left(\frac{1}{2\varepsilon}\sqrt{\frac{4\varepsilon k^2 - \sigma^2 \mu}{\mu}}\right)t\right]\exp\left[-\frac{\sigma}{2\varepsilon}t\right]$, which is the source-free solution and where we can assume that the wavevector has the direction of any of the Cartesian unit vectors. This mode along with the identical one propagating in the opposite direction can be used as an expansion basis to form the general solution. Additionally, losses effect the field in time since the fields occupy the whole space as assumed in Section II (contradictory to the stationary-medium plane-wave equivalent). The analytic signal analysis gives us the complex solutions of **U** ($\tilde{\mathbf{B}}$ or $\tilde{\mathbf{D}}$):

$$\mathbf{U} = \begin{cases} \mathbf{U}_0 e^{\mathbf{i}\mathbf{k}\cdot\mathbf{r} - \left(\frac{\sigma_1}{2\varepsilon_1} + \frac{\mathrm{i}}{2\varepsilon_1}\sqrt{\frac{4\varepsilon_1 k^2 - \sigma_1^2 \mu_1}{\mu_1}}\right)t}, & t < 0 \\ \mathbf{U}_0 S_{11} e^{\mathbf{i}\mathbf{k}\cdot\mathbf{r} - \left(\frac{\sigma_2}{2\varepsilon_2} + \frac{\mathrm{i}}{2\varepsilon_2}\sqrt{\frac{4\varepsilon_2 k^2 - \sigma_2^2 \mu_2}{\mu_2}}\right)t} \\ +\mathbf{U}_0 S_{21} e^{\mathbf{i}\mathbf{k}\cdot\mathbf{r} - \left(\frac{\sigma_2}{2\varepsilon_2} - \frac{\mathrm{i}}{2\varepsilon_2}\sqrt{\frac{4\varepsilon_2 k^2 - \sigma_2^2 \mu_2}{\mu_2}}\right)t}. & t > 0 \end{cases} \quad (16)$$

From the continuity relations we get: $S_{11} = \frac{1}{2}\left(\frac{\sqrt{\varepsilon_1\mu_1} + \sqrt{\varepsilon_2\mu_2}}{\sqrt{\varepsilon_1\mu_1}} + \mathrm{i}\frac{(\sigma_2\varepsilon_1 - \sigma_1\varepsilon_2)}{2\varepsilon_1 k}\sqrt{\frac{\mu_2}{\varepsilon_2}}\right)$ and $S_{21} = \frac{1}{2}\left(\frac{\sqrt{\varepsilon_1\mu_1} - \sqrt{\varepsilon_2\mu_2}}{\sqrt{\varepsilon_1\mu_1}} - \mathrm{i}\frac{(\sigma_2\varepsilon_1 - \sigma_1\varepsilon_2)}{2\varepsilon_1 k}\sqrt{\frac{\mu_2}{\varepsilon_2}}\right)$. The complete temporal scattering matrix of Eq. (15) can be easily found analogously:

$$S_0 = \begin{pmatrix} \frac{1}{2}\left(\frac{\sqrt{\varepsilon_1\mu_1} + \sqrt{\varepsilon_2\mu_2}}{\sqrt{\varepsilon_1\mu_1}} + \mathrm{i}\frac{(\sigma_2\varepsilon_1 - \sigma_1\varepsilon_2)}{2\varepsilon_1 k}\sqrt{\frac{\mu_2}{\varepsilon_2}}\right), & \frac{1}{2}\left(\frac{\sqrt{\varepsilon_1\mu_1} - \sqrt{\varepsilon_2\mu_2}}{\sqrt{\varepsilon_1\mu_1}} + \mathrm{i}\frac{(\sigma_2\varepsilon_1 - \sigma_1\varepsilon_2)}{2\varepsilon_1 k}\sqrt{\frac{\mu_2}{\varepsilon_2}}\right) \\ \frac{1}{2}\left(\frac{\sqrt{\varepsilon_1\mu_1} - \sqrt{\varepsilon_2\mu_2}}{\sqrt{\varepsilon_1\mu_1}} - \mathrm{i}\frac{(\sigma_2\varepsilon_1 - \sigma_1\varepsilon_2)}{2\varepsilon_1 k}\sqrt{\frac{\mu_2}{\varepsilon_2}}\right), & \frac{1}{2}\left(\frac{\sqrt{\varepsilon_1\mu_1} + \sqrt{\varepsilon_2\mu_2}}{\sqrt{\varepsilon_1\mu_1}} - \mathrm{i}\frac{(\sigma_2\varepsilon_1 - \sigma_1\varepsilon_2)}{2\varepsilon_1 k}\sqrt{\frac{\mu_2}{\varepsilon_2}}\right) \end{pmatrix}. \quad (17)$$

This scattering matrix has interesting properties. We observe that: $S_{11} + S_{21} = 1$, $S_{12} + S_{22} = 1$, $\det(S_0) = \sqrt{\frac{\varepsilon_1\mu_1}{\varepsilon_2\mu_2}}$ and the inverse $S_0^{-1}$ corresponds to the scattering by the time-reversed material transition (namely from $\varepsilon_2, \mu_2, \sigma_2$ to $\varepsilon_1, \mu_1, \sigma_1$). In addition, if $\eta = \sqrt{\mu/\varepsilon}$ and $\sigma/\varepsilon$ remain constant $S_{21}$ and $S_{12}$ extinguish, i.e. the wave does not reflect. The scattering matrix formalism can be equally used for standing waves without neccesarily reformulating the complex solutions of $\mathbf{R}(\mathbf{r})$, $T(t)$ since standing waves are a superposition of propagating solutions of opposite direction.

Taking into account temporal dispersion [19], high frequency field components may be generated. However these generated amplitudes with these frequencies are expected insignificant, and exponentially decaying with time (due to the conservation of momentum).

### III. EXCEPTIONAL POINTS AND OPERATOR SYMMETRIES ANALYSIS

The electromagnetic modes and the resulting parametric amplification from time-periodic media have been extensively studied by [6], [8] using the solutions of the corresponding hypergeometric differential equations [25], [26]. In this section, we consider the simple case of a temporal material jump, where Maxwell's equations permit the discontinuity of the **E** and **H** fields and temporal translational symmetry is broken (as first studied in [1]) and connect the parametric amplification phenomena with the exceptional-point analysis, that is often used in Hamiltonian physics, without assuming time-periodic modulation. In addition, we examine this time-domain scattering problem under parity and time-reversal operators offering extra physical insights to this wave problem.

#### A. Exceptional points

In Hamiltonian physics, an exceptional point is an operating condition, in which (at least) two eigenvalues and eigenmodes collide and the effective Hamiltonian looses (at least) one of its dimensions and exhibits thus a degeneracy [27]. Such theory can be extended to time-varying systems by formulating a state matrix, which models the time-transitioning behavior of the problem [28]. Let us assume that at $t = \tau$ there is an abrupt change in the medium from $\varepsilon_1, \mu_1, \sigma_1$ to $\varepsilon_2, \mu_2, \sigma_2$ that can be modeled by the scattering formulation of the previous Section (under the appropriate conditions). We define the phase transmission matrix:

$$\mathbf{T}_\kappa(t) = \begin{pmatrix} e^{\left(-\frac{\sigma_\kappa}{2\varepsilon_\kappa} - \frac{\mathrm{i}}{2\varepsilon_\kappa}\sqrt{\frac{4\varepsilon_\kappa k^2 - \sigma_\kappa^2 \mu_\kappa}{\mu_\kappa}}\right)t} & 0 \\ 0 & e^{\left(-\frac{\sigma_\kappa}{2\varepsilon_\kappa} + \frac{\mathrm{i}}{2\varepsilon_\kappa}\sqrt{\frac{4\varepsilon_\kappa k^2 - \sigma_\kappa^2 \mu_\kappa}{\mu_\kappa}}\right)t} \end{pmatrix}, \quad (18)$$

where $\kappa = 1, 2$ and the matrix of coefficients is $\mathbf{C}(t) = \begin{bmatrix} C_1 & C_2 \end{bmatrix}^T$, where $C_1$ and $C_2$ are the amplitudes of $T_1(t)$ and $T_2(t)$ respectively [as shown in Eqs. (13)-(14)] for a fixed value of the wavenumber. The matrix of coefficients is:

$$\mathbf{C}(\tau') = \mathbf{T}_1(\tau) S_0^{-1} \mathbf{T}_2(\tau' - \tau) S_0 \mathbf{C}(0) = \hat{H}(\tau')\mathbf{C}(0). \quad (19)$$

The time-transitioning state matrix of this system is $\hat{H}(\tau') = \mathbf{T}_1(\tau) S_0^{-1} \mathbf{T}_2(\tau' - \tau) S_0$, with $\det(\hat{H}(\tau')) = \exp\left[-\left(\frac{\sigma_1}{\varepsilon_1}\tau + \frac{\sigma_2}{\varepsilon_2}(\tau'-\tau)\right)\right]$, where $\tau'$ is the time of observation after the abrupt change of material properties ($\tau' > \tau$). If we consider a lossless medium $\det(\hat{H}(\tau')) = 1$ then its corresponding eigenvalues and eigenmodes are:

$$\lambda_\pm = b \pm \sqrt{b^2 - 1} \qquad (20)$$

$$\mathbf{C}_+ = \begin{pmatrix} a \\ \lambda_+ + c \end{pmatrix} \qquad (21)$$

$$\mathbf{C}_- = \begin{pmatrix} a \\ \lambda_- + c \end{pmatrix} \qquad (22)$$

where:

$$b(k,\varepsilon_1,\mu_1,\varepsilon_2,\mu_2,\tau,\tau') = \cos\left(\frac{k}{\sqrt{\varepsilon_1\mu_1}}\tau\right)\cos\left(\frac{k}{\sqrt{\varepsilon_2\mu_2}}(\tau'-\tau)\right)$$
$$- \frac{(\varepsilon_1\mu_1 + \varepsilon_2\mu_2)}{2\sqrt{\varepsilon_1\mu_1\varepsilon_2\mu_2}}\sin\left(\frac{k}{\sqrt{\varepsilon_1\mu_1}}\tau\right)\sin\left(\frac{k}{\sqrt{\varepsilon_2\mu_2}}(\tau'-\tau)\right), \qquad (23)$$

and $a = -\frac{i(\varepsilon_1\mu_1 - \varepsilon_2\mu_2)}{2\sqrt{\varepsilon_1\mu_1\varepsilon_2\mu_2}}\sin\left(\frac{k}{\sqrt{\varepsilon_2\mu_2}}(\tau'-\tau)\right)\exp\left(-i\frac{k}{\sqrt{\varepsilon_1\mu_1}}\tau\right)$,

$c = \exp\left(-i\frac{k}{\sqrt{\varepsilon_1\mu_1}}\tau\right)\left[-\cos\left(\frac{k}{\sqrt{\varepsilon_2\mu_2}}(\tau'-\tau)\right) + \frac{i(\varepsilon_1\mu_1 + \varepsilon_2\mu_2)}{2\sqrt{\varepsilon_1\mu_1\varepsilon_2\mu_2}}\sin\left(\frac{k}{\sqrt{\varepsilon_2\mu_2}}(\tau'-\tau)\right)\right]$. $b$ depends on multiple parameters as we observe from Eq. (23) and determines the energy transitions from the time-dynamic material and field interactions. The factor $b$ determines the energy conversions that occur as the wave takes energy from the modulation or the modulation takes energy from the wave and their overall energy-transition characteristics. The detected exceptional points are connected with the marginal stability conditions of the hypergeometric equations which correspond to the associated parametrically amplifying modes in time-periodic media and Hamiltonians [6], [8], [28]–[30].

The eigenvalues satisfy the condition: $\lambda_+\lambda_- = 1$. The system supports two propagating modes when $|b| < 1$ (as both eigenvalues are complex and unitary), and one evanescent and one parametrically amplifying for $|b| > 1$ (as one is lower and the other higher than unity). The special transition case: $|b| = 1$ is associated with exceptional points since the eigenvalues and the eigenmodes collide: $\lambda_+ = \lambda_- = \pm 1$ and $\mathbf{C}_+ = \mathbf{C}_-$.

A plot of the eigenvalue transitions is given in Fig. 1. In Fig. 1(a) the real part of the eigenvalues is shown as a function of the parameter $b$. In Fig. 1(b) the position of the eigenvalues on the complex plane are presented (arrows showing the position of the eigenvalues as $b$ changes values from a negative value -3 to a positive value +3, as a combination of changes of its variables).

As proven in [1], a step temporal material jump results in a gain ratio of the electromagnetic energy density: $u_2/u_1 = (\varepsilon_1\mu_2 + \mu_1\varepsilon_2)/(2\varepsilon_2\mu_2)$, where $u = \varepsilon|\mathbf{E}|^2$, but this ratio is not the only quantitative value which defines the energy gain. Indeed the energy interactions depend also on the time variables [6], [8]. As presented in the stability charts of Fig. 2, where we fix $k/\sqrt{\varepsilon_1\mu_1}$, $k/\sqrt{\varepsilon_2\mu_2}$ and vary the time variables: $\tau$ and $\tau'$, for a selected group of values of the time variables, the medium supports evanescent and amplified modes, despite that the material properties are unchanged. If $k/\sqrt{\varepsilon_1\mu_1}$ swaps value with $k/\sqrt{\varepsilon_2\mu_2}$, then $\tau \leftrightarrow \tau'-\tau$. This property of the stability charts indicates certain operator symmetries in time and space, which are studied in the next subsection.

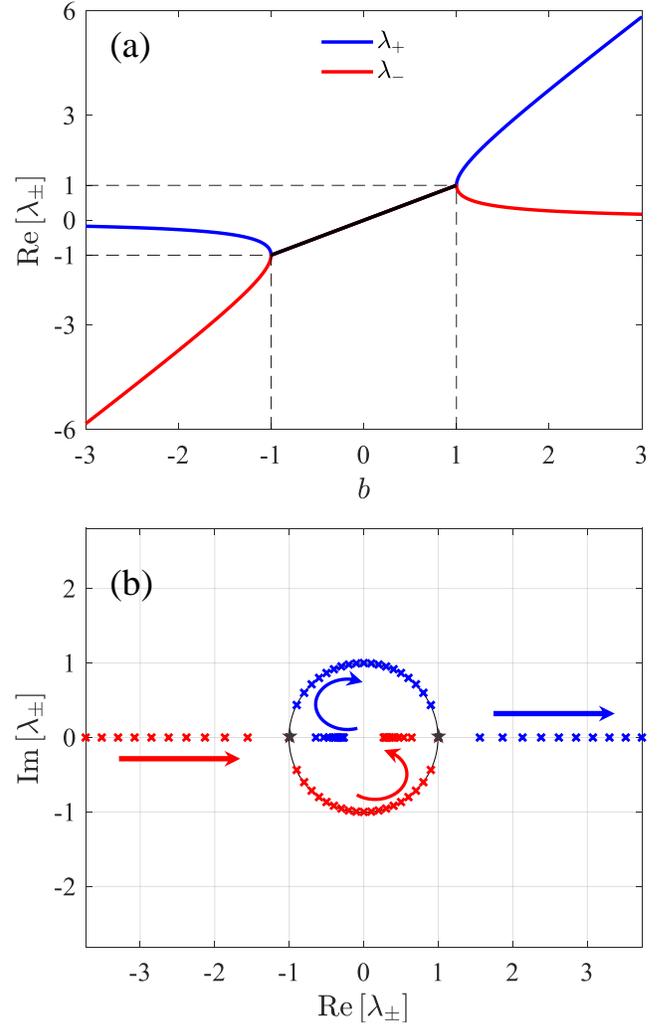

Fig. 1. Graphical representations of the transitions of the eigenvalues of the time-transitioning state matrix of a system in which an abrupt change of material parameters occurs. (a) Evolution of the real part of the eigenvalues and (b) chart showing both real and imaginary part of the eigenvalues as $b$ shifts values. For $|b| < 1$ the eigenvalues are complex and the wave-states are stable, for $|b| > 1$ the eigenvalues become real representing amplification, while at $|b| = 1$ we observe the exceptional points ($\lambda_+ = \lambda_- = \pm 1$). Arrows at (b) show the quantitative change of the eigenvalues as $b$ increases in value (from -3 to +3).

### B. Parity, time-reversal and parity-time symmetries

The effect of the parity operator on the matrix of coefficients is $\mathcal{P}\mathbf{r} = -\mathbf{r}$. We can define the parity operator as:

$$\mathcal{P}\mathbf{C}(t) = \mathcal{P}\begin{pmatrix} C_1 \\ C_2 \end{pmatrix} = \begin{pmatrix} C_2^* \\ C_1^* \end{pmatrix}. \qquad (24)$$

We obtain for real $\lambda_\pm$: $\mathcal{P}\mathbf{C}_+ = \mathbf{C}_+$ and $\mathcal{P}\mathbf{C}_- = \mathbf{C}_-$. Whereas for complex $\lambda_\pm$: $\mathcal{P}\mathbf{C}_+ = \mathbf{C}_-$ and $\mathcal{P}\mathbf{C}_- = \mathbf{C}_+$. The physical interpretation of these results is very interesting. In the case of real eigenvalues, the change of direction is not going to affect the amplified or evanescent electromagnetic modes. On the contrary, in the case of the complex eigenvalues, a change in the direction is very important as the direction plays an important role (shown also in a time-periodic medium by solving the Hill equation in [6]).

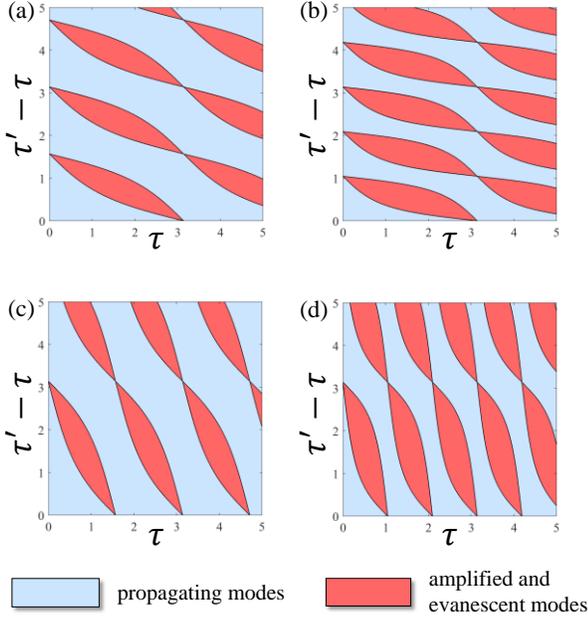

Fig. 2. Stability charts for fixed values of $k/\sqrt{\varepsilon_1\mu_1}$ and $k/\sqrt{\varepsilon_2\mu_2}$: (a) $k/\sqrt{\varepsilon_1\mu_1} = 1$, $k/\sqrt{\varepsilon_2\mu_2} = 2$, (b) $k/\sqrt{\varepsilon_1\mu_1} = 1$, $k/\sqrt{\varepsilon_2\mu_2} = 3$, (c) $k/\sqrt{\varepsilon_1\mu_1} = 2$, $\frac{k}{\sqrt{\varepsilon_2\mu_2}} = 1$ and (d) $k/\sqrt{\varepsilon_1\mu_1} = 3$, $k/\sqrt{\varepsilon_2\mu_2} = 1$.

The time-reversal operator indicates that $\mathcal{T}t = -t$, is defined thus as:

$$\mathcal{T}\mathbf{C}(t) = \mathcal{T}\begin{pmatrix} C_1 \\ C_2 \end{pmatrix} = \begin{pmatrix} C_2 \\ C_1 \end{pmatrix}. \quad (25)$$

For time-varying homogeneous media, time-reversal is not the complex conjugate operator as it would have been if the problem was stationary [16]. The time-reversal relation of the state matrix is:

$$\left(\mathcal{T}\hat{H}^{-1}(\tau')\mathcal{T}\right)\left(\mathcal{T}\mathbf{C}_\pm\right) = \lambda_\pm^{-1}\left(\mathcal{T}\mathbf{C}_\pm\right), \quad (26)$$

where $\mathcal{T}\mathbf{C}_\pm$ are the new eigenvectors and $\mathcal{T}\hat{H}^{-1}(\tau')\mathcal{T}$ is the new state matrix. If the modulation is symmetric, meaning that starting from $t = 0$ we observe the waves at a time: $\tau' = 2\tau$ where the medium has spent equal time having $\varepsilon_1, \mu_1$ and $\varepsilon_2, \mu_2$ then $\mathcal{T}\hat{H}^{-1}(\tau')\mathcal{T}$ has interchanged eigenvalues and eigenvectors with $\hat{H}(\tau')$. This result can be physically interpretated by imagining that we could run time backwards. In that case, the amplified mode would look as if it attenuates and the evanescent one would look like it is amplified.

This analysis for parity and time-reversal operators shows that $\mathcal{PT}$ – symmetry is achievable as a double interchange of eigenvalues and eigenmodes overtly brings back the system to its initial form. Such $\mathcal{PT}$ – symmetric wave-states are exhibited without the presence of material gain and loss. The only necessary condition is a symmetric time-modulation of a dynamic medium [29].

IV. CONCLUSIONS

In this paper, we revised, reexamined and expanded the solutions derived in [1] for the electromagnetic field theory of a background time-modulated medium. We assumed time-dependent and non-dispersive $\varepsilon(t)$, $\mu(t)$, $\sigma(t)$ and derive the time-domain solutions of the fields based on the variable separation technique, the system's causality and the momentum-Fourier transforms (in contrast with the nomenclature in electromagnetics where it is convenient to work in the frequency domain). The momentum of the fields is conserved, while energy is not and we provided the appropriate theory to deal with temporal discontinuities. At last, we connected the intriguing energy transition phenomena that occur with the exceptional point theory used in Hamiltonian physics, examined the effects of parity and time-reversal operators and proved that parity-time symmetry can be present in a symmetrically time-modulated and time-aperiodic medium without a symmetrically combined distribution of gain and loss. For the latter observation the analysis was performed for an abrupt change of medium parameters assuming an instantaneous-responding medium with instantaneous time variation of the signals. Such assumptions may divert from an experimental demonstration since time-dynamic local materials may not be realistic or easily implementable. Nevertheless, the overall analysis may be useful and enriches the existing literature on time-varying systems and the topic of electromagnetic wave propagation under nonstationary media.

APPENDIX A: NONSTATIONARY AND INHOMOGENEOUS MEDIUM EQUATIONS

The electromagnetic equations for $\varepsilon = \varepsilon(\mathbf{r},t)$, $\sigma = \sigma(\mathbf{r},t)$ and $\mu = \mu(\mathbf{r},t)$ are:

$$\nabla \times \left(\frac{1}{\mu}\right)\nabla \times \mathbf{E} + \varepsilon\frac{\partial^2 \mathbf{E}}{\partial t^2} + \left[\sigma + 2\frac{\partial \varepsilon}{\partial t} + \varepsilon\frac{\partial \ln \mu}{\partial t}\right]\frac{\partial \mathbf{E}}{\partial t} + \left[\frac{\partial \sigma}{\partial t} + \frac{\partial^2 \varepsilon}{\partial t^2} + \left(\frac{\partial \varepsilon}{\partial t} + \sigma\right)\frac{\partial \ln \mu}{\partial t}\right]\mathbf{E} = \mathbf{H} \times \nabla\left(\frac{\partial \ln \mu}{\partial t}\right) - \left(\frac{\partial \ln \mu}{\partial t} + \frac{\partial}{\partial t}\right)\mathbf{J}, \quad (27)$$

$$\nabla \times \left(\frac{1}{\varepsilon}\right)\nabla \times \mathbf{H} + \mu\frac{\partial^2 \mathbf{H}}{\partial t^2} + \left[2\frac{\partial \mu}{\partial t} + \left(\frac{\sigma}{\varepsilon} + \frac{\partial \ln \varepsilon}{\partial t}\right)\mu\right]\frac{\partial \mathbf{H}}{\partial t} + \left[\frac{\partial^2 \mu}{\partial t^2} + \left(\frac{\sigma}{\varepsilon} + \frac{\partial \ln \varepsilon}{\partial t}\right)\frac{\partial \mu}{\partial t}\right]\mathbf{H} = -\mathbf{E} \times \nabla\left(\frac{\sigma}{\varepsilon} + \frac{\partial \ln \varepsilon}{\partial t}\right) + \nabla \times \left(\frac{1}{\varepsilon}\right)\mathbf{J}. \quad (28)$$

Eqs. (27)-(28) describe the electrodynamic-wave interactive phenomena with material time-variation and inhomogeneities. In contrast with a regular inhomogeneous medium [20], the differential equations have added terms associated with the unusual properties imparted by the time-variation in the derivatives and in the sources. Not only does the current radiate differently in comparison with a stationary medium, but also an extra pseudo-source appears related to an unusual field coupling; such is physically associated with equivalent charge distributions, which depend on both the inhomogeneity and the time-variation [3], [5].

A sufficient condition for any general medium to involve only $\mathbf{E}$ for (27) is: $\nabla\left(\frac{\partial \ln \mu}{\partial t}\right) = 0$. Such condition is naturally satisfied in stationary media. This condition implies, after elementary algebraic manipulation, that $\mu$ is a product of a function that depends only on space, $\mu_r$ and a function that depends only on time, $\mu_t$: $\mu = \mu_r \mu_t$. The electric field equation (27) becomes (in the absence of sources):

$$\nabla \times \mu_r^{-1} \nabla \times \mathbf{E} + \frac{\partial}{\partial t}\left[\mu_t \left(\frac{\partial(\varepsilon\mathbf{E})}{\partial t} + \sigma\mathbf{E}\right)\right] = 0. \quad (29)$$

For the magnetic field equation, the sufficient condition to decouple the Eq. (28) and involve only $\mathbf{H}$ is: $\nabla\left(\frac{\sigma}{\varepsilon} + \frac{\partial \ln \varepsilon}{\partial t}\right) = 0$. This condition is not naturally satisfied for a time-invariant medium, because the conductivity of the medium induces electric-displacement currents even for a stationary problem. Analogously with the electric field case the condition implies that that $\varepsilon$ is a function of the form: $\varepsilon = \varepsilon_r \varepsilon_t \exp\left(-\int_\tau^t \frac{\sigma}{\varepsilon} dt\right)$, for $t \geq \tau$, where $\varepsilon_r$ only depends on space and $\varepsilon_t$ only depends on time. In the special case that the value $\sigma/\varepsilon$ is only time-dependent: $\sigma = \sigma_r \sigma_t$ with $\varepsilon_r = \sigma_r$. The magnetic field equation (28) becomes (in the absence of sources):

$$\nabla \times \varepsilon_r^{-1} \nabla \times \mathbf{H} + \frac{\partial}{\partial t}\left[\varepsilon_t e^{-\int_\tau^t \frac{\sigma}{\varepsilon} dt} \frac{\partial(\mu\mathbf{H})}{\partial t}\right] + \sigma_t \frac{\partial(\mu\mathbf{H})}{\partial t} = 0. \quad (30)$$

In cases of space-time periodic media, the solutions are found using the Bloch-Floquet ansatz [31-36].

APPENDIX B: POYNTING AND MOMENTUM CONSERVATION THEOREMS

The conservation of energy takes the form [37]:

$$\frac{\partial W}{\partial t} + \nabla \cdot \mathbf{S} = -p, \quad (31)$$

where $W$ is the total stored energy, $\mathbf{S}$ the Poynting vector and $p$ is the power supplied by an external agent. In the case of a homogeneous time-varying medium:

$$W = \frac{\varepsilon(t)}{2}|\mathbf{E}|^2 + \frac{\mu(t)}{2}|\mathbf{H}|^2, \quad (32)$$

$$\mathbf{S} = \mathbf{E} \times \mathbf{H}, \quad (33)$$

$$p = \mathbf{E} \cdot (\mathbf{J} + \sigma(t)\mathbf{E}) + \frac{1}{2}\frac{d\mu(t)}{dt}|\mathbf{H}|^2 + \frac{1}{2}\frac{d\varepsilon(t)}{dt}|\mathbf{E}|^2. \quad (34)$$

It is evident that $W$ and $\mathbf{S}$ remain the same as in any regular electromagnetic problem, but $p$ has extra terms which depend on the time-derivative of $\varepsilon, \mu$. The redefined $p$ power factor can give us a quantitative proof that time-varying media have different energy-transfer characteristics and could possibly allow for amplification.

The momentum conservation theorem takes the form [37]:

$$\frac{\partial \mathbf{G}}{\partial t} + \nabla \cdot \mathbb{T} = -\mathbf{f}, \quad (35)$$

where $\mathbf{G}$ is the momentum density vector, $\mathbb{T}$ is the Maxwell stress tensor and $\mathbf{f} = \rho\mathbf{E} + (\mathbf{J} + \sigma(t)\mathbf{E}) \times \mathbf{B}$ is the Lorentz force. It is straightforward to show that under a time-varying homogeneous medium the momentum conservation theorem remains the same as in the stationary case, namely:

$$\mathbb{T} = \varepsilon(t)\mathbf{EE} + \mu(t)\mathbf{HH} - \frac{1}{2}\left(\varepsilon(t)|\mathbf{E}|^2 + \mu(t)|\mathbf{H}|^2\right)\mathbb{I}, \quad (36)$$

$$\mathbf{G} = \varepsilon(t)\mu(t)\mathbf{E} \times \mathbf{H}. \quad (37)$$

The electromagnetic field momentum is invariant under homogeneous time-variations of the medium (in classical wave theory this means that the wavenumber is conserved) whereas its energy as shown from the Poynting theorem analysis alters (which also indications a frequency shift, since from a quantum-mechanics point of view the energy is proportional to the frequency).

APPENDIX C: POTENTIALS, LORENTZ GAUGE AND HERTZ VECTOR

The electric and magnetic fields can be represented by the scalar $V$ and the $\mathbf{A}$ potentials:

$$\mathbf{E} = -\nabla V - \frac{\partial \mathbf{A}}{\partial t}, \quad \mathbf{B} = \nabla \times \mathbf{A}. \quad (38)$$

The two corresponding equations for the potentials in homogeneous time-varying media are:

$$\nabla \cdot (\varepsilon(t)\nabla V) + \nabla \cdot \left(\varepsilon(t)\frac{\partial \mathbf{A}}{\partial t}\right) = -\rho,$$

$$\nabla \times \nabla \times \mathbf{A} + \mu(t)\frac{\partial(\varepsilon(t)\nabla V)}{\partial t} + \mu(t)\frac{\partial(\varepsilon(t)\mathbf{A})}{\partial t} \quad (39)$$

$$+ \mu(t)\sigma(t)\nabla V + \mu(t)\sigma(t)\frac{\partial \mathbf{A}}{\partial t} = \mu(t)\mathbf{J}.$$

Since the potentials are not unique, we can choose a gauge so that the equations are decoupled. The gauge condition that may be used is:

$$\nabla \cdot \mathbf{A} + \mu(t)\frac{\partial(\varepsilon(t)V)}{\partial t} + \mu(t)\sigma(t)V = 0. \quad (40)$$

This gauge fixing resembles the Lorentz gauge condition for stationary media. Under this condition the decoupled equation is:

$$\nabla^2 \mathbf{A} - \mu(t)\frac{\partial}{\partial t}\left(\varepsilon(t)\frac{\partial \mathbf{A}}{\partial t}\right) - \mu(t)\sigma(t)\frac{\partial \mathbf{A}}{\partial t} = -\mu(t)\mathbf{J}. \quad (41)$$

Notice that Eqs. (6) and (41) share the same homogeneous solutions. Since we added the gauge condition the unknowns are three identities, we introduce thus the electric Hertz vector $\mathbf{\Pi}$: $\mathbf{A} = -\mu(t)\partial\mathbf{\Pi}/\partial t$ and $V = \nabla \cdot \mathbf{\Pi}/\varepsilon(t)$.

$$\nabla^2 \mathbf{\Pi} - \varepsilon(t)\frac{\partial}{\partial t}\left(\mu(t)\frac{\partial \mathbf{\Pi}}{\partial t}\right) - \int \sigma(t)\frac{\partial}{\partial t}\left(\mu(t)\frac{\partial \mathbf{\Pi}}{\partial t}\right)dt = \int \mathbf{J} dt, \quad (42)$$

where the electric and magnetic fields are: $\mathbf{H} = -\nabla \times \partial\mathbf{\Pi}/\partial t$ and $\mathbf{E} = -\nabla\nabla \cdot \mathbf{\Pi}/\varepsilon(t) + \frac{\partial}{\partial t}\left(\mu(t)\frac{\partial}{\partial t}\mathbf{\Pi}\right)$.

APPENDIX D: SECOND ORDER DIFFERENTIAL EQUATIONS OF $T(k,t)$ AND THE LIOUVILLE-GREEN APPROXIMATION

The differential equation of $T(k,t)$ is:

$$\frac{d^2 T(k,t)}{dt^2} + P(t)\frac{dT(k,t)}{dt} + Q(k,t)T(k,t) = 0, \quad (43)$$

where for **B**: $P(t) = \frac{\sigma}{\varepsilon} + \frac{d\ln\varepsilon}{dt}$, $Q(k,t) = \frac{k^2}{\varepsilon\mu}$ and for **D**: $P(t) = \frac{\sigma}{\varepsilon} + \frac{d\ln\mu}{dt}$ and $Q(k,t) = \frac{1}{\mu}\frac{d}{dt}\left(\frac{\mu\sigma}{\varepsilon}\right) + \frac{k^2}{\varepsilon\mu}$. If $x(k,t)$ is the homogeneous solution of the magnetic induction differential equation then the homogeneous solution $y(k,t)$ for the electric displacement differential equation is: $y(k,t) = \exp\left(-\int\gamma dt\right)\int x(k,t)\left[\exp\left(\int\gamma dt\right)\right]dt$, where $\gamma = \sigma/\varepsilon$. The Liouville-Green estimation is [22]:

$$T(k,t) \approx \frac{\sqrt{\exp\left(-\int P(t)dt\right)}}{\sqrt[4]{Q(k,t)}}\exp\left(\pm\int\sqrt{-Q(k,t)}dt\right). \quad (44)$$

Eq. (44) is applied directly for the magnetic induction $T_b(k,t)$ (as shown in Eq. (8)), whereas the relation between $x(k,t)$ and $y(k,t)$ and Eq. (8) are used to derive (9).

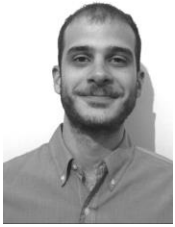
**Theodoros T. Koutserimpas** received the Diploma degree in electrical and computer engineering from the National Technical University of Athens, Athens, Greece, in 2017. He is currently pursuing the Ph.D. degree in electrical engineering with the Swiss Federal Institute of Technology in Lausanne, Lausanne, Switzerland. His research interests include, but are not limited to, electrodynamics, computational electromagnetics, microwaves, acoustics, optics and applying novel numerical and analytical methods to problems dealing with wave phenomena.

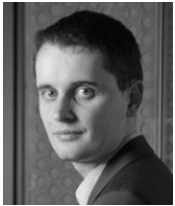
**Romain Fleury** is a tenure-track assistant professor in the Institute of Electrical Engineering at the Swiss Federal Institute of Technology in Lausanne (EPFL), and currently leads the EPFL Laboratory of Wave Engineering. He received the M.S. degree in micro and nanotechnology from the University of Lille, Lille, in 2010, and the Ph.D. degree in electrical and computer engineering from the University of Texas at Austin, Austin, TX, USA, in 2015. In 2016, he was a Marie-Curie postdoctoral fellow at ESPCI Paris-Tech, and CNRS Langevin Institute, Paris, France. His research interests include a wide variety of topics in the field of wave physics and engineering, including periodic structures, active and time-modulated metamaterials, nonreciprocal wave propagation, classical topological insulators, nonlinear effects, and parity-time symmetry. He has published more than 30 articles in peer-reviewed scientific journals, including papers in journals such as Science, Physical Review Letters, and Nature Communications. His work on nonreciprocal acoustics was featured on the cover of science, and attracted the attention of the general public, with appearances in various media including NBC News, Daily Mail, and Scientific American. In 2014, he received the Best Student Paper Award in Engineering Acoustics as well as the Young Presenter Award in Noise from the Acoustical Society of America. His research on parity-time symmetric acoustics has been awarded the Best Student Paper at the International Congress Metamaterials in 2014. He also received the F.V. Hunt award from the Acoustical Society of America, and several other awards including the 2016 Outstanding Reviewer Award from Institute of Physics Publishing. He is passionate about teaching and is responsible for two bachelor courses on Electromagnetic field theory in the Institute of Electrical Engineering at EPFL.